\documentclass{article}
\usepackage{cite}
\usepackage{graphicx}
\usepackage{float}

\def\bq{\begin{eqnarray}}
\def\eq{\end{eqnarray}}
\def\cir{\phi}

\begin{document}

\renewcommand*{\thefootnote}{\Alph{footnote}}
\setcounter{footnote}{1}

\title{\bf Bouncing Dirac particles: compatibility between
MIT boundary conditions and Thomas precession}

\author {Nistor Nicolaevici\footnote{email: \texttt{nnicolaevici@gmail.com}}
\\
\it Department of Physics, The West University of Timi\c soara,
\\
\it V. P\^arvan 4, 300223, Timi\c soara, Romania}

\maketitle

\begin{abstract}

We consider the reflection of a Dirac plane wave on a perfectly reflecting plane described by chiral
MIT boundary conditions and determine the rotation of the spin in the reflected component of the wave.
We solve the analogous problem for a classical particle using the evolution of the spin defined by the
Thomas precession and make a comparison with the quantum result. We find that the rotation axes
of the spin in the two problems coincide only for a vanishing chiral angle, in which case the rotation
angles coincide in the nonrelativistic limit,\, and also remain remarkably close  in the relativistic regime.
The result shows that in the nonrelativistic limit the interaction between the spin and a reflecting
surface with nonchiral boundary conditions  is completely contained in the Thomas precession
effect, in conformity with the  fact that these boundary  conditions are equivalent to an infinite repulsive scalar
potential outside the boundary. By contrast, in the ultrarelativistic limit the rotation angle in the quantum
problem remains finite, while in the classical one the rotation angle diverges. We comment on the possible
implications of this discrepancy on the validity of the Mathisson-Papapetrou-Dixon equations at large
accelerations.

\end{abstract}

\renewcommand{\thefootnote}{\arabic{footnote}}


\section*{1 Introduction}

One of the interesting aspects of the Dirac equation is the connection with the dynamics of a classical
spinning particle. Since the Dirac equation describes a particle with spin, such a connection
is expected in the classical limit of the Dirac theory. For a nonrelativistic particle in a weak
electromagnetic field, this is a well-known subject \cite{bjor, itzy}. The connection is usually
realized by a succession of Foldy-Wouthuysen (FW) transformations, which projects
the Hamiltonian on a Pauli two-spinor, from which the classical limit can be easily read. Some years
ago, the result was extended also to relativistic particles in strong electromagnetic fields
\cite{sile}. Using a more sophisticated FW transformation, it was shown there
that the evolution equations for the Heisenberg operators in the FW representation  with
$\hbar \rightarrow 0$ exactly reproduce the corresponding classical equations of motion for a particle with spin.
In particular, the evolution of the quantum spin reduces to the classical evolution defined by the
Bargman-Michel-Telegdi equation \cite{barg}.

A similar situation appears to be valid for a test particle in the gravitational field.
The generally accepted equations of motion in this case are the Mathisson-Papapetrou-Dixon\footnote{For
a history of the subject and recent developments see ref.~\cite{putz}.} (MPD) equations \cite{math, papa, dixo}.
The fact that the MPD equations can be extracted from the classical limit of the Dirac equation
(neglecting terms quadratic in spin) has been established using various methods
by many authors \cite{wong,kane,cate,audr,baru,cian1,cian2,obuk1,obuk2}, with the observation
that the connection is by no means unique. A main source of ambiguity for a particle in a curved
background is the nonunique definition of the center of mass and implicitly of the spin of the
particle, which is intimately linked to the freedom of choice of the supplemental condition which
has to be imposed on the spin tensor in order to close the  MPD equations \cite{costa}. It seems
that there is yet no general agreement on which the ``correct'' supplemental conditions are, and
whether different supplemental conditions lead or not to equivalent physical pictures \cite{costa, stei},
and the semiclassical limit of the Dirac equation might shed a light on these questions \cite{cian1, obuk2}.

The intention of this paper is to investigate the quantum-classical connection in the Dirac theory, by considering
a concrete problem in the usual flat spacetime. The problem is as follows: a particle bounces off from a perfectly
reflecting plane. It is clear that the picture in the orbital space has nothing special, so that the interesting
part is the evolution of the spin. Our aim is to determine the rotation of the spin after the reflection from the
plane in the quantum and the classical description, and make a comparison between the two results. We are especially
interested in this problem due to the following reason. In the quantum description, we will consider a family of
boundary conditions on the reflecting plane. In the classical description we will consider the simple case in which there
is no specific interaction between the spin and the plane. It is then a good question which boundary conditions are
the best fit for the classical result. Although this is a rather simple problem, it seems that it has not been
investigated yet. We will see that some notable features emerge. We stress in advance that in both descriptions
the solution can be exactly obtained, which will allow a perfectly accurate comparison between the quantum and
the classical picture.

We begin with an outline of the calculation. In the quantum description, the key ingredient are the
boundary conditions to be imposed on the perfectly reflecting plane. The common choices in literature for
the Dirac field are the MIT \cite{chod, john, dono} and the spectral \cite{atiy, horta, falom} boundary
conditions. We will use the MIT boundary conditions, which are the closest ones to the usual Dirichlet or
Neumann conditions for the scalar field. (In both cases the normal component of the density current on the
boundary is required to vanish.) It will be also relevant for the comparison with the classical picture
to consider the extension of the MIT conditions \cite{lutk, theb} involving a chiral angle $\phi$.
We recall that the effect of a nonvanishing chiral angle is to introduce an additional interaction between
the field and the boundary, and that this extra interaction can significantly change the physics of the
system. (A well-known example are the chiral bag models for nucleons \cite{theb, brow, vento, hosa}; see also below.)

Two important observations for our discussion are as follows. First, the nonchiral case $\phi=0$ corresponding
to the original MIT boundary condition is the precise equivalent of an infinite repulsive scalar potential
outside the region  enclosed\footnote{Or, equivalently, an infinite mass of the field in this region.} by the
reflecting boundary \cite{chod}. It is then immediate from the scalar nature of the potential that in this case
there is no spin-dependent interaction between the particle and the boundary. By contrast,
a nonvanishing angle $\phi\neq 0$ generally introduces a specific interaction between the
boundary and the spin \cite{hosa}. A simple example which  illustrates this fact is provided by the free Dirac
particle in the exterior of a perfectly reflecting sphere with chiral MIT conditions on the surface \cite{jaffef}. It turns
out for $\phi=0$ only scattering states exist, but for $\phi \neq 0$ a finite number of bound states are also
allowed. These bound states can be naturally interpreted in terms of an attraction force between the sphere and
the particle. More specifically, with an appropriate chiral transformation one can show that a nonzero angle
$\phi\neq 0$ is equivalent to the nonchiral MIT boundary condition, plus a delta-type magnetic field
localized on the surface of the sphere. The attraction force between the particle and the chiral sphere can then be
understood as a consequence of the interaction between the particle's spin (magnetic moment) and this effective
magnetic field \cite{jaffef}.

Returning to our goal, the calculation of the orientation of the spin in the reflected component of the
wave is directly similar to that for the polarization of an electromagnetic plane wave scattered by a
perfectly reflecting plane. The only essential difference is that we now have to use the plane wave
solutions of the Dirac equation. As expected, the MIT boundary conditions plus the  kinematic constraints in
the orbital space completely determine the relation between the incident and the reflected spin. This relation
can be expressed in terms of a SU(2) rotation matrix, from which we will extract the
rotation axis and rotation angle of the spin.

We now turn to the classical problem. We will make the key assumption that the interaction between the
particle and the reflective plane is such that no external torque acts on the particle. As a consequence,
the spin four-vector is Fermi-Walker transported along the trajectory. We will consider the evolution of
the three-spin in the proper frame of the particle, with the particle's proper frame defined via a Lorentz
boost with respect to the laboratory frame. In these conditions, the Fermi-Walker transport implies that the
spin evoluates according to the Thomas precession formula \cite{whee, jack}. It is clear that on the inertial parts of
the trajectories, $i.e.$ before and after the impact with the plane, the spin remains fixed. The nontrivial
part is what happens at the impact point. A first observation is that, assuming a perfectly reflecting plane,
the velocity is non-differentiable at this point, which is rather unphysical. We will remediate this situation
by considering that the reflecting plane acts via a repulsive potential, smoothing thus the trajectory. The
evolution of the spin will then be determined by the Thomas precession due to the accelerated
motion in this potential. Let us assume that the reflecting plane coincides with the $xy$-plane.
We will naturally choose the repulsive potential to be of the form $V(z)$, assuring thus the translational
invariance of the\, system along the plane. At first sight, one would expect the orientation
of the spin after reflection to depend on the form of the potential  $V(z)$. Notably, we will find
that this is not so. Hence, the comparison with the quantum picture will not be affected by the
ambiguity in the choice of the repulsive potential associated to the plane.

In brief,\, the main conclusions are as follows. Not surprisingly, we will find that the
closest similarity between the quantum and the classical picture arises only for a chiral angle
$\phi=0$.\footnote{As we will see there is actually a second possibility $\phi=\pi$, but
this is equivalent to the case $\phi=0$.} For this particular value of $\phi$ the rotation axes
of the spin in the two cases coincide, and for nonrelativistic velocities of the incident particle the
rotation angles also coincide. Moreover, the two rotation angles remain remarkably close up to velocities
comparable with the speed of light. These show that for a vanishing chiral angle and not too high velocities
the evolution of  the spin at the impact with the plane is completely described by the Thomas precession
effect. This is precisely what one would expect from the  fact that the nonchiral MIT conditions correspond to
a scalar potential associated to the surface, in which case no specific spin-dependent interaction  exist.

By contrast, in the ultrarelativistic limit a rather unexpected discrepancy shows up, and this is perhaps
the most interesting result of the paper: the  rotation angle of the spin in the quantum picture remains
finite, while in the classical picture the rotation angle diverges. The divergent angle  in the classical problem 
seems to be unnatural, which will invite us to speculate that something goes wrong in the classical calculation.
We will suggest that a more realistic model for a spinning particle which includes the elasticity of the body
would leave the rotation angle in the ultrarelativistic limit finite. Generalizing, we will argue that a
similar problem could appear for the MPD equations at large accelerations of the particle.

The paper is organized as follows. In sect.~2 we determine the rotation of the spin in the quantum
problem. In sect.~3 we obtain the analogous quantity in the classical problem, and in sect.~4 we compare
the two results. We end in sect.~5 by presenting the conclusions and the possible implications for the MPD equations.
Throughout the paper we use natural units with $\hbar=c=1$.

\section*{2 The quantum rotation angle}

The first step is to obtain the wave function in the presence of the reflecting plane. As usual, the result
can be written as a superposition of an incident and a reflected component. For a definite momentum of the
particle, these components are plane waves solutions\, of the Dirac equation. It is useful to first recall the
form of these solutions. We naturally restrict to  the positive energy solutions.

We work in the standard Dirac representation in which the gamma matrices are ($\sigma_i$ are the Pauli matrices)
\bq
\qquad
\gamma^0 =
\left|
\begin{array}{cc}
I & 0
\\
0 & -I
\end{array}
\right|,
\quad
\gamma^i =
\left|
\begin{array}{cc}
0 & \sigma_i
\\
-\sigma_i & 0
\end{array}
\right|, \quad i =1,2,3.
\eq
The positive energy plane wave solutions\, can be written as\, (standard notation is used)
\bq
u_{p,\, \xi}(x)= u(p, \xi) \,e^{-ip\cdot x},
\label{usol}
\eq
where $p^\mu=(E, {\bf p})$ is the four-momentum of the particle and $\xi$ a unit normalized two-spinor
which defines the spin. The four-spinors $u(p, \xi)$ can be obtained as follows. One first introduces
the spinors in the proper frame of the particle corresponding to the four-momentum $\hat p^\mu= (m, {\bf 0})$, $i.e.$
\bq
\qquad\qquad
u(\hat p,\, \xi)=
\left|
\begin{array}{c}
\xi
\\
0
\end{array}
\right|, \quad\xi^+\xi=1.
\eq
The spin in the particle's proper frame is then defined by $\xi$ the same as in the
nonrelativistic theory. The four-spinors $u(p, \xi)$  can be obtained as
\bq
u(p,\, \xi)= S(\Lambda_p)\, u(\hat p,\, \xi),
\eq
where $\Lambda_p$ is a Lorentz transformation which sends $\hat p$ into $p$. We choose it as usual
to be a pure Lorentz boost,
\bq
S(\Lambda_p)=e^{-i\alpha {\bf n} \cdot {\bf K}}, \quad K_i=\frac{i}{2}\,\gamma_i \gamma_0,
\eq
where
\bq
\alpha =\mbox{arctanh}\frac{\vert {\bf p}\vert}{E}, \quad {\bf n}=\frac{\bf p}{ \vert {\bf p}\vert}.
\eq
The result is \cite{bjor, itzy}:
\bq
u(p,\, \xi)=
\left|
\begin{array}{c}
\cosh \frac{\alpha}{2}\,\xi
\\
\sinh \frac{\alpha}{2}\, \sigma_{\bf n}\, \xi
\end{array}
\right|,
\quad \sigma_{\bf n} \equiv {\bf n} \cdot \mbox{\boldmath $\sigma$}.
\eq
By fixing the Lorentz  transformations $\Lambda_p$  the proper frame of the particle is completely determined,
making thus precise the significance of $\xi$. Note that the same proper frame of the particle will be used in the classical
calculation, which will make immediate the comparison with the quantum result.

We now discuss the boundary conditions on the reflecting plane. Let us introduce the four-vector
\bq
N^\mu=(0, {\bf N}),
\eq
where ${\bf N}$ is the unit normal to the plane. The MIT conditions with a chiral angle $\phi$ are defined
by \cite{lutk, theb}
\bq
N^\mu \gamma_\mu \psi = i e^ {i\cir \gamma^5} \psi,
\label{mitboc}
\eq
with  $\gamma^5 \equiv i\gamma^0\gamma^1\gamma^2\gamma^3$. We recall that in the right member
\bq
e^ {i\cir \gamma^5}=\cos \cir\, I +i\sin \cir \gamma_5,
\label{gamex}
\eq
and that in the Dirac representation
\bq
\gamma^5
=\left|
\begin{array}{cc}
0 & I
\\
I & 0
\end{array}
\right|.
\eq
In the case of interest the reflecting surface is the $xy$-plane, and thus\footnote{The sign
ambiguity in the normal ${\bf N}$ can be included in $\phi\rightarrow \phi \pm\pi$.}
\bq
{\bf N}= {\bf e}_z.
\label{ortvec}
\eq
Let us  decompose the Dirac wave function as
\bq
\psi=
\left|
\begin{array}{c}
\Phi
\\
\chi
\end{array}
\right|.
\label{desc}
\eq
Combining the relations above one finds that condition (\ref{mitboc}) translates into
\bq
(\sigma_3 +\sin \cir)\, \Phi-i\cos \cir\, \chi=0,
\label{c1}
\\
(\sigma_3-\sin \cir)\, \chi+i\cos \cir\, \Phi=0.\,
\label{c2}
\eq
One can easily check  that eqs. (\ref{c1}) and (\ref{c2}) are linearly dependent, as it should,
in order to admit a non-zero solution. It is therefore sufficient to consider only one of the two
equations.

Let us choose the directions of the incident $(I)$ and reflected $(R)$ directions as (see fig.~1)
\bq
{\bf n}_I=(\cos \theta,\, 0, -\sin\theta),
\label{m4}
\quad
{\bf n}_R=(\cos \theta,\, 0,\,\sin\theta).
\eq
\begin{figure}[!]

\centerline{\includegraphics[width=2.7in, height=4.5in, angle=90]{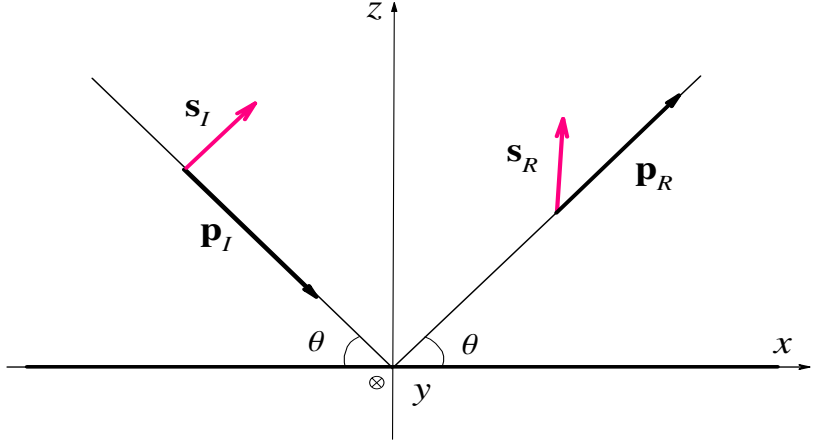}}

\caption{The geometry of the problem.}

\end{figure}
The corresponding momenta are
\bq
p_{\,I/R}^{\,\mu}=m\, (\cosh \alpha,\, \sinh \alpha\, {\bf n}_{I/R}).
\eq
The total wave function can then be written as
\bq
\psi(x)=u(p_{\,I},\, \xi_I)\, e^{-i p_{\,I}\cdot x}+
u(p_R,\, \xi_R)\, e^{-i p_R\cdot x},
\label{wavef}
\eq
where the two four-spinors are
\bq
u(p_I,\, \xi_I)=
\left|
\begin{array}{c}
\cosh \frac{\alpha}{2}\,\xi_I
\\
\sinh \frac{\alpha}{2}\, \sigma_{I}\,\xi_I
\end{array}
\right|,
\quad
\sigma_{I} = {\bf n}_{I}\! \cdot  \mbox{\boldmath $\sigma$}
,\,\,
\eq
\bq
u(p_R,\, \xi_R)=
\left|
\begin{array}{c}
\cosh \frac{\alpha}{2}\,\xi_R
\\
\sinh \frac{\alpha}{2}\, \sigma_{R}\,\xi_R
\end{array}
\right|,
\quad
\sigma_{R} = {\bf n}_{R} \cdot \mbox{\boldmath $\sigma$}.
\eq
It is clear that the spinors $\xi_I$ and $\xi_R$ define the orientations of the incident and the reflected spin.

We are interested in the relation between $\xi_I$ and $\xi_R$. Applying  one of the conditions (\ref{c1}) or
(\ref{c2}) to the wave function (\ref{wavef}) on the plane $z=0$, a simple calculation shows that the result
can be put in the following form:
\bq
Q_I\, \xi_I= Q_R\, \xi_R,
\label{qir}
\eq
where $Q_I$ and $Q_R$ are the matrices
\bq
Q_I=AI + {\bf B} \cdot \mbox{\boldmath $\sigma$},
\quad
Q_R=A^*I + {\bf B} \cdot \mbox{\boldmath $\sigma$},
\eq
with $A$ and ${\bf B}$ given by
\bq
A=1+i\cos \cir \sin \theta \tanh\frac{\alpha}{2},\qquad\quad\,\,
\label{A}
\\
{\bf B} = \cos \cir \cos \theta \tanh \frac{\alpha}{2}\,{\bf e}_y + \sin \cir \,{\bf e} _z.\,
\label{B}
\eq
Note that $A$ is complex and ${\bf B}$ is real. Introducing the new matrix
\bq
{\cal U}=Q_{R}^{-1} Q_I,
\label{uqq}
\eq
eq. (\ref{qir}) becomes
\bq
\xi_R= {\cal U}\, \xi_I.
\eq
Using the obvious properties
\bq
Q_I^+=Q_R, \quad [Q_I, Q_I^+]=0,
\eq
it is easy to show that ${\cal U}$ is a unitary matrix
\bq
{\cal U}^+{\cal U}=I.
\label{unit}
\eq
This means that the matrix ${\cal U}$ can be readily interpreted as a rotation operator in the spin space.

Let  us denote by  $\varphi_q$  and ${\bf n}_q$ the rotation angle and, respectively, the (unit-norm) rotation axis
implied by ${\cal U}$. We can write
\bq
{\cal U}= e^{i\chi}\left ( \cos\frac {\varphi_q}{2}\, I- i \sin \frac{\varphi_q}{2}\, {\bf n}_q\cdot
\mbox{\boldmath $\sigma$}
\right),
\label{su2}
\eq
where $e^{i\chi}$ is a pure phase and where in the parentheses one recognizes the standard form of an SU(2)
matrix in the angle-axis parameterization. Constructing the matrix ${\cal U}$\, from eqs. (\ref{qir})-(\ref{uqq})
and comparing with eq. (\ref{su2}), one finally finds:
\bq
\cos\frac {\varphi_q}{2}=\frac
{\vert A\vert^ 2 - {\bf B}^{\,2}}
{\vert A^2 - {\bf B}^{\,2}\vert},
\quad
\sin\frac {\varphi_q}{2}=\frac
{2\, \mbox{Im}A\,\vert \bf B\vert}
{\vert A^2 - {\bf  B}^{\,2}\vert},
\quad
{\bf n}_q =\frac{\bf B}{\vert \bf B\vert}.
\label{qpack}
\eq
This practically solves the quantum problem. Note that $\varphi_q$ and ${\bf n}_q$ are completely determined by three
parameters: the velocity parameter $\alpha$, the incident angle $\theta$ and the chiral angle $\phi$ (the mass $m$ of
the particle is irrelevant). The expressions (\ref{qpack}) in the general case are somewhat complicated and we will
not write them down in explicit form. A significant simplification will occur when we will make contact with the
classical picture in sect.~4.

\section*{3 The classical rotation angle}

We first recall some basic facts about the evolution of the classical spin. Let us denote the spin four-vector of the particle
by $S^\mu(\tau)$, with $\tau$ the proper time along the trajectory. Assuming that no external torque acts on the particle, the
evolution of the spin is given by the Fermi-Walker transport
\bq
\frac{dS^\mu}{d\tau}=\Omega^\mu_{\,\,\nu} \, S^\nu,
\quad \Omega^\mu_{\,\,\nu}=u^\mu a_\nu - a^\mu u _\nu,
\eq
where $u^\mu$ and $a^\nu$ are the four-velocity and  four-acceleration of the particle.
Using the Lorentz transformations $\Lambda_p$ introduced in the previous section, the spin four-vector in
the particle's proper frame
is
\bq
\qquad\hat S^\mu(\tau)=[\Lambda_{p\,(\tau)}^{-1}]^\mu_{\,\,\,\nu} S^\nu(\tau), \quad
p\, (\tau)=m u(\tau).
\eq
The orthogonality requirement between the four-spin and the four-velocity
\bq
S^\mu u_\mu=0,
\eq
implies that
$\hat S^\mu$ is of the form
\bq
\hat S^\mu= (0, {\bf s}\,),
\eq
where the three-vector ${\bf s}$ can be identified with the usual spin measured in the particle's  proper frame. It is
clear that with these definitions ${\bf s}$\, is the direct correspondent of the spin determined by the spinor $\xi$
in the quantum problem.

As is well known, the evolution of the spin $\bf s$\, is defined by the Thomas precession \cite{whee, jack}.
\bq
\frac{d{\bf s}}{dt}=\mbox{\boldmath $\omega$}_T\times{\bf s},
\quad
\mbox{\boldmath $\omega$}_T=\frac{\gamma^2}{\gamma+1}\, {\bf a} \times {\bf v},
\label{thopre}
\eq
where $\bf v$ and $\bf a$ are the classical velocity and acceleration of the particle, and $\gamma$ is the usual
relativistic factor.\footnote{The Thomas precession effect remains a constant subject of investigation. For some
recent works on the effect see $e.g.$ refs. \cite{t1, t2, t3, t4}.} We now apply eqs. (\ref{thopre}) to determine the relation
between the incident and the reflected spin in the problem of interest.

On the inertial parts of the trajectory, when the particle is not in contact with the plane, the acceleration is
${\bf a} =0$, which implies $\mbox{\boldmath $\omega$}_T=0$, and thus ${\bf s}\,=\mbox{constant}$.
The nontrivial evolution of the spin is determined by the interaction with the plane. As discussed in sect.~1, the idea
is to describe this interaction via a region of repulsive potential $V(z)$, where we will choose the potential
to be zero for $z>0$. The picture in this case looks like in fig.~2. We are interested in the rotation of the
spin after passing through the repulsive region $z\leq 0$. We will make the simplificatory assumption that the
spin-orbit
coupling $E_{\footnotesize\mbox{s-o}}={\bf s} \cdot \mbox{\boldmath $\omega$}_T$ is sufficiently small compared
to the potential energy $E_{\footnotesize\mbox{pot}}=V(z)$, so that the trajectory in the orbital space is
practically the same with that of a particle without spin.\, This would correspond to a sufficiently small spin
${\bf s}$\, and/or a sufficiently large mass $m$ of the particle, which will keep the acceleration\, ${\bf a}$
small, and thus the precession $\mbox{\boldmath $\omega$}_T$ small.

\begin{figure}[h]

\centerline{\includegraphics[width=2.7in, height=4.5in, angle=90]{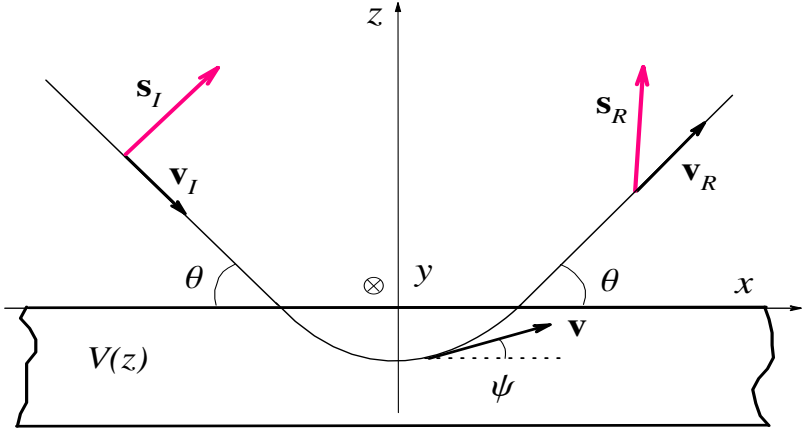}}

\caption{The classical trajectory in the presence of the reflecting surface modeled by a repulsive potential
in the region  $z\leq0 $.}

\end{figure}

It is clear from the symmetries of the problem that we can consider that the trajectory is contained in the $xz$-plane.
Let  us denote the by $\psi$  the angle between the velocity ${\bf v}$ and the $x$-axis.  One can read
from fig.~2 that for the piece of the trajectory $z\leq 0$ this angle varies within the interval
\bq
\psi \in [-\theta, \theta]. \label{tlim}
\eq
The trick is to parameterize the velocity in the following form:
\bq
{\bf v}(\psi)=v(\psi)\, {\bf e}(\psi), \quad {\bf e} (\psi)=(\cos \psi, 0,\, \sin \psi),\label{vel}
\eq
where $\psi$ has to be seen as a function of the laboratory time $t$. The acceleration of the particle
then is
\bq
{\bf a}(\psi) = \frac{d{\bf v}}{dt}=\left (\frac{d v}{d\psi}\, {\bf e}(\psi)+ v(\psi) \frac{d  {\bf e}}{d\psi}\right)\,
\frac{d\psi}{dt}. \label{acc}
\eq
Introducing eqs. (\ref{vel}) and (\ref{acc}) in the Thomas precession formula (\ref{thopre}), one finds that the
precession vector is (in obvious notation)
\bq
\mbox{\boldmath $\omega$}_T=
\frac{d\varphi_{c}}{dt}\, {\bf n}_{\,c},
\eq
where the angular velocity and the rotation axis are
\bq
\frac{d\varphi_{c}}{dt} =\frac{\gamma^2v^2}{\gamma+1}\, \frac{d \psi}{dt},
\quad {\bf n}_{\,c}= {\bf e}_y.
\label{clares}
\eq
Note that, essentially, the rotation axis is independent of time. In these conditions the total rotation angle of the
spin $\varphi_{c}$ can be simply obtained by integrating the angular velocity $d\varphi_{c}/dt$ with respect to $t$.
Due to the derivative factor $d\psi/dt$ in the angular velocity,  the integration can be readily
replaced by that with respect to $\psi$ with the integration limits (\ref{tlim}). One thus obtains
\bq
\varphi_{c}=\int_{-\theta}^{\,\theta} d\psi \frac{\gamma^2v^2}{\gamma+1},
\label{priint}
\eq
where it remains to find the dependence $v=v(\psi)$. This can be easily done by noting that the translational invariance
along the plane ensures the conservation of the momentum along the axis $x$, which implies
\bq
\gamma (\psi) v(\psi) \cos \psi = \gamma  v  \cos \theta,
\label{con}
\eq
where in the right member are the quantities before/after on the impact on the plane. Using eqs.
(\ref{priint}) and (\ref{con}) a few manipulations lead to
\bq
\varphi_{c}= (\gamma  v \cos \theta)^2 \times
\int_{-\theta}^{\,\theta} \frac{d \psi}{\cos^2\psi}
\left(\sqrt{1+\frac{(\gamma v \cos \theta)^2}{\cos^2\psi}}+1\right)^{-1}.
\label{clarot}
\eq
This solves the classical problem. Thus, the picture is that the spin rotates around the
axis\, ${\bf e}_y$ with the angle $\varphi_{c}$ defined by eq. (\ref{clarot}). As anticipated, the
rotation angle is independent of the form of the potential $V(z)$. We will explore some consequences of our
results in the following section.

\section*{4 Comparison between the two results}

We are interested in the values of the chiral angle $\phi$ which lead to the closest
similarity between the evolution of the spin in the quantum and the classical picture. We begin
by looking at the rotation axes. From eqs. (\ref{qpack}) and (\ref{B}) the axis in the quantum problem is
\bq
{\bf n}_{q}\sim \cos \cir \cos \theta \tanh \frac{\alpha}{2}\,{\bf e}_y + \sin \cir \,{\bf e} _z.
\label{qapr}
\eq
We have seen that in the classical problem
${\bf n}_{\,c}={\bf e}_y.$
Comparing with eq. (\ref{qapr}) it is immediate that the two axes coincide only when $\sin \phi =0$, $i.e.$
\bq
\phi = 0\,\,\, \mbox{or}\,\,\pi.
\eq
A simple calculation using eqs. (\ref{A}), (\ref{B}) and (\ref{qpack}) shows that the case $\phi=0$ implies
\bq
\label{phi0}
\tan \frac{\varphi_q}{2}=\frac
{\sin 2\theta \,\tanh^2 (\alpha/2)}{1-\cos 2\theta\, \tanh^2(\alpha/2)},
\quad
{\bf n}_q={\bf e}_y.
\eq
Using the same formulas one finds that the case $\phi=\pi$ is equivalent to replacing in the expressions
above
\bq
\varphi_q \rightarrow -\varphi_q,
\quad
{\bf n}_q \rightarrow -{\bf n}_q.
\eq
This brings nothing new,\, as it defines the same rotation\footnote{This means that for $\phi=0$ the sign
ambiguity in the normal ${\bf N}$ in the boundary condition (\ref{mitboc}) is irrelevant. The same property
appears in other results \cite{gold, zahe}. The property however is not universal, as illustrated by the Casimir
energies between two  plates \cite{bene} or inside a cylinder \cite{ambr} with MIT chiral boundary
conditions.} as eq. (\ref{phi0}). In the rest of the section, we will therefore continue to refer only to
the case $\phi=0$.

We now compare the rotation angles. We begin by considering the nonrelativistic limit $v \ll 1$. In this
limit we can approximate in eq. (\ref{phi0})
\bq
\alpha \simeq v, \quad \tanh (\alpha/2) \simeq v/2,
\eq
and since $\tanh^2 (\alpha/2)\ll 1$ the denominator of the long fraction can be approximated to unity. In
these conditions the quantum angle becomes
\bq
\varphi^{NR}_q \simeq \frac{1}{2}\, v^2 \sin 2\theta.
\label{quaner}
\eq
In the classical angle (\ref{clarot}) the relativistic factor is $\gamma \simeq 1$, and using $v^2\ll 1$ the inverse
round brackets under the integral can be approximated to $1/2$. This leads to
\bq
\Delta \phi^{NR}_{c} \simeq
(v \cos \theta)^2 \times
\int_{-\theta}^{\,\theta} \frac{d \psi}{2\cos^2\psi},
\nonumber
\\
=\frac{1}{2}\, v^2 \sin 2\theta.
\qquad\qquad\qquad\,\,
\label{uf}
\eq
The conclusion from eqs. (\ref{quaner}) and (\ref{uf}) is that in the nonrelativistic limit the quantum and the
classical rotation angles coincide.

\begin{figure}

\centerline{\includegraphics[width=4.6in, height=5.9in, angle=90]{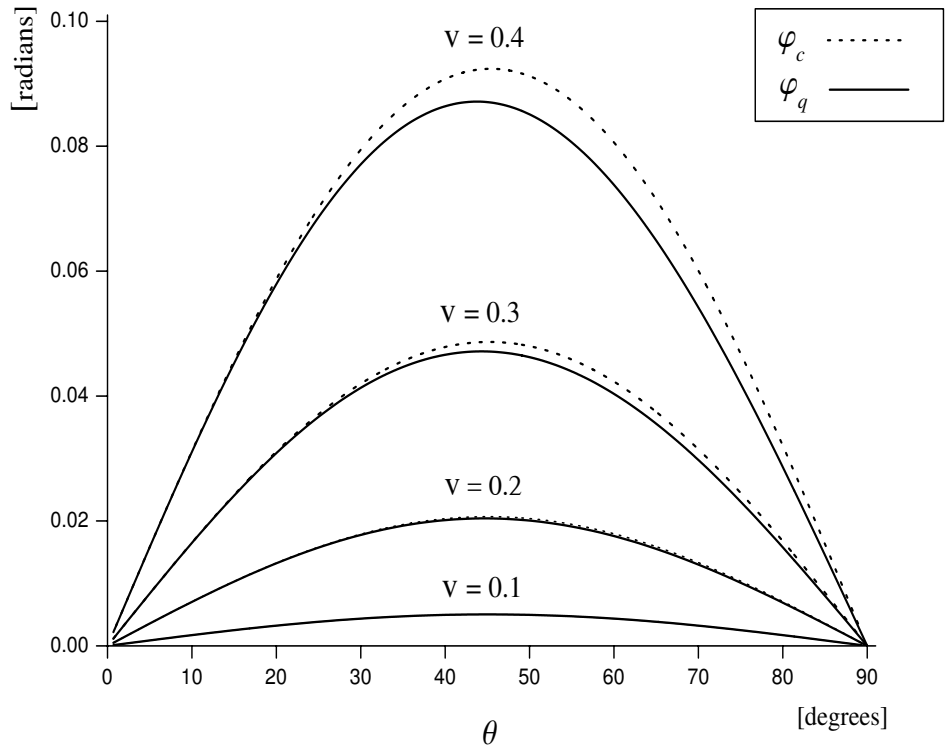}}

\caption{The rotation a angles of the spin in the quantum (solid line) and the classical
problem (dotted line) shown as a function of the angle $\theta$ for a number of velocities
$v$ not too close to the speed of light $c=1$. For velocities smaller than $v\sim 0.1$ the curves
are indistinguishable on the plot.}

\end{figure}

It is interesting to see what happens at larger velocities $v$. In fig.~3 we represented the rotation angles
as a function of\, $\theta$\, for different values of $v$. The curves show that the two angles are practically
indistinguishable for velocities up to $v\sim 0.1$. Remarkably, the two angles remain fairly equal also
for relativistic velocities as high as $v\sim 1/2$.

However, as $v$ increases the classical angle tends to assume larger values and the curves become
significantly different. A plot for a highly relativistic velocity $v=0.9$ is shown in fig.~4.

\begin{figure}[h]

\centerline{\includegraphics[width=4.4in, height=5.8in, angle=90]{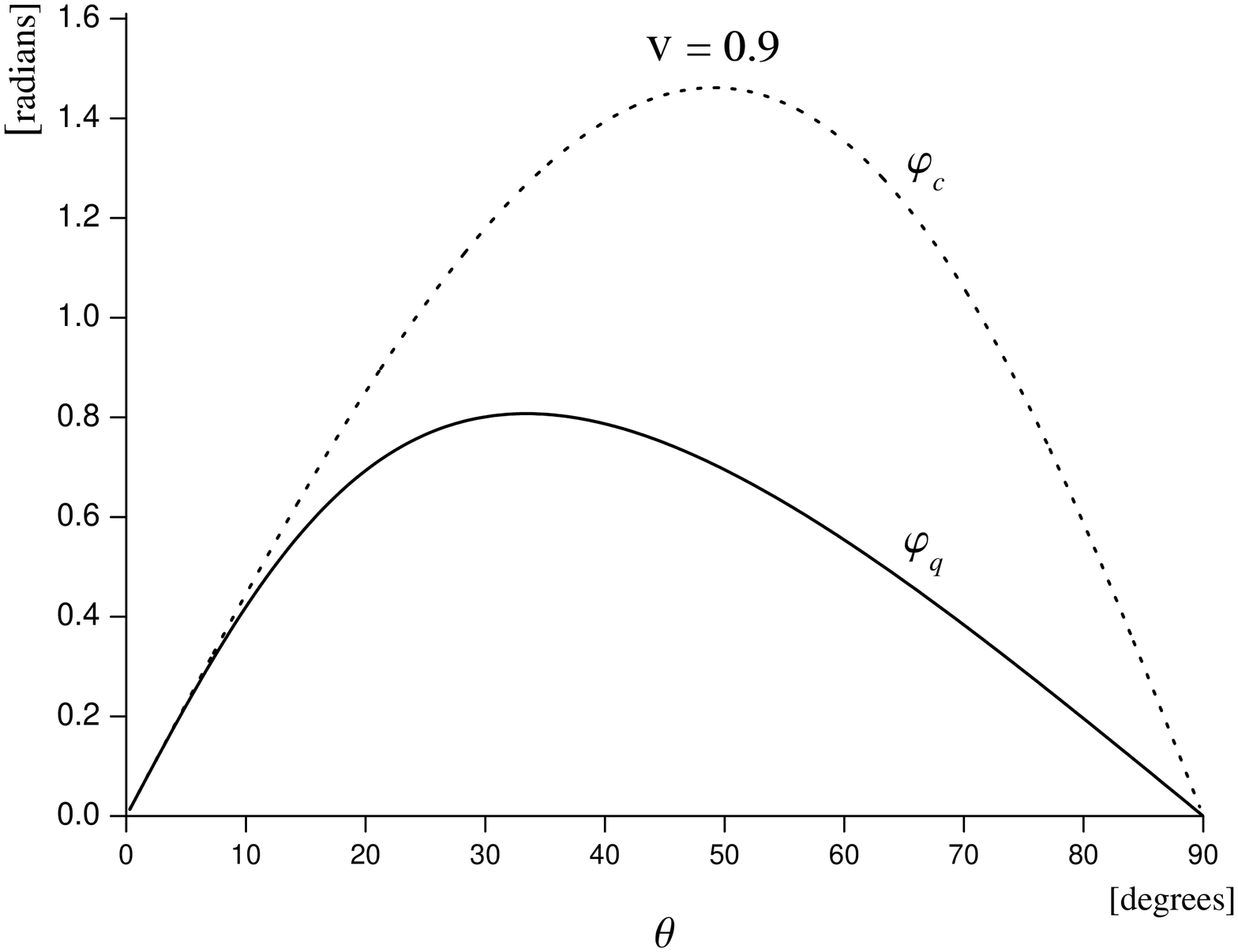}}

\caption{The same as in fig.~1 but for $v=0.9$. For highly relativistic velocities the classical rotation
angles generally become significantly larger than the quantum ones.}

\end{figure}

We now consider the ultrarelativistic limit $v\rightarrow 1$. In this case in the quantum angle (\ref{phi0})
$\tanh \alpha/2\rightarrow 1$, from which
\bq
\tan \frac{\phi^{\,UR}_q}{2}\simeq
\frac {\sin 2\theta} {1-\cos 2\theta},
\eq
which is equivalent to
\bq
\varphi_q^{UR}\simeq \pi - 2\theta.
\label{rotuvl}
\eq
Note that the result remains finite. We will give a simple geometric interpretation for eq. (\ref{rotuvl}) a few
lines below. A plot which shows  $\varphi_q$ as a function of the velocity parameter $\alpha$ for various
angles $\theta$ is given in fig.~5.

\begin{figure}[h]

\centerline{\includegraphics[width=4.4in, height=5.8in, angle=90]{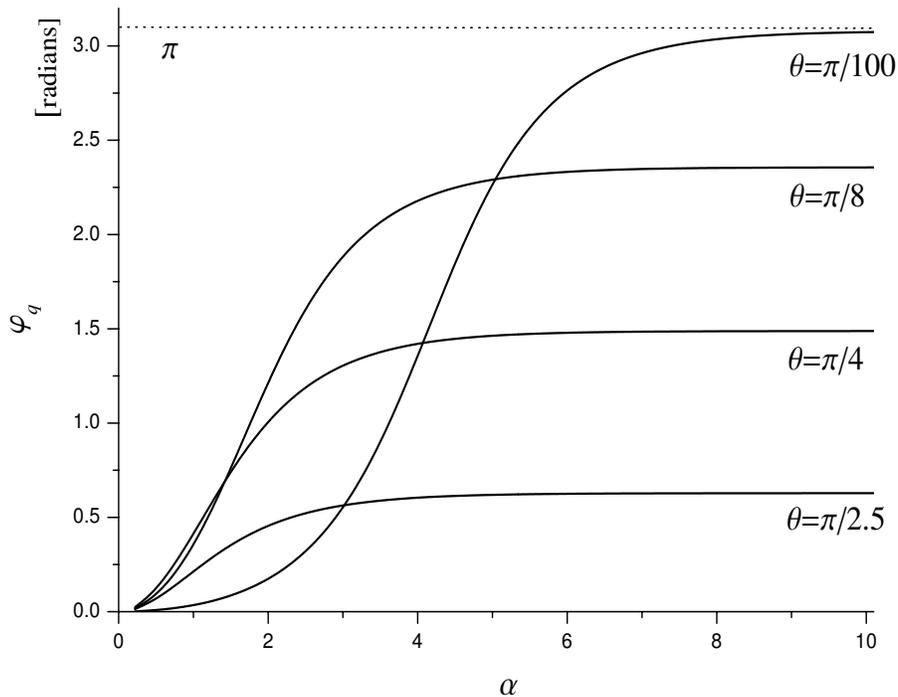}}

\caption{The quantum rotation angle $\varphi_q$ as a function of the velocity parameter $\alpha$
for  different angles $\theta$. The horizontal pieces of the curves at large $\alpha$ correspond to
the ultrarelativistic limit (\ref{rotuvl}).}

\end{figure}

A significantly different result is obtained in the classical picture. In the ultrarelativistic limit
$\gamma\rightarrow \infty$, and counting the powers of  $\gamma$ in eq. (\ref{clarot}) one can see that
that $\varphi_c$ behaves as
\bq
\varphi_c^{UR} \sim \gamma \rightarrow \infty.
\eq
This means that in the ultrarelativistic limit the rotation angle  diverges ($i.e.$, the
spin makes an infinite number of turns). A plot which illustrates the divergent behavior of $\varphi_c$
compared to that of $\varphi_q$ is shown in fig.~6. We will comment on the discrepancy between the
quantum and the classical result in the next section.

\begin{figure}[h]

\centerline{\includegraphics[width=4.2in, height=5.6in, angle=90]{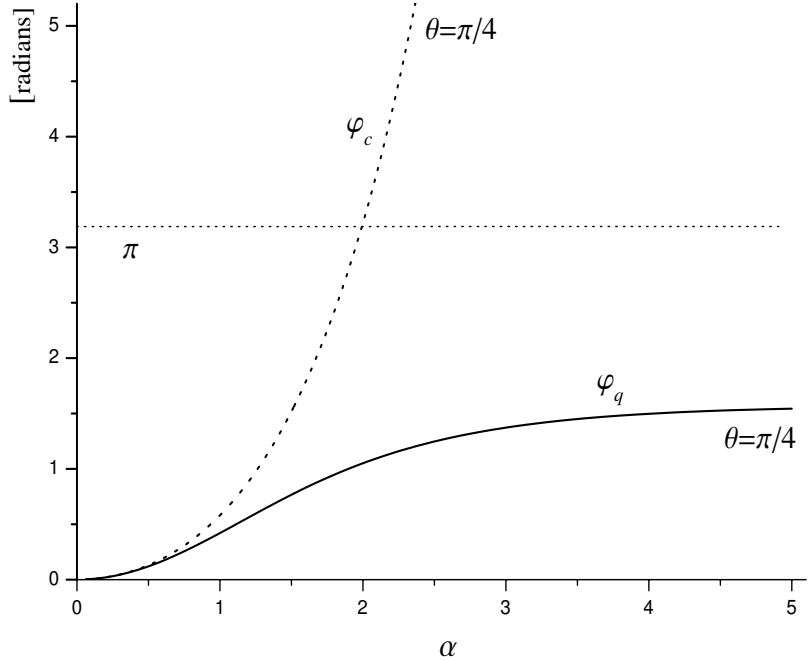}}

\caption{The quantum and the classical rotation angles as functions of $\alpha$ for a
fixed incidence angle $\theta$. In the ultrarelativistic limit $\varphi_q$ remains finite,
while $\varphi_c$ diverges.}

\end{figure}

Finally, although not directly related to our subject, it deserves to observe that eq. (\ref{rotuvl})
admits the following interpretation. Note first that, after the reflection from the plane, the momentum ${\bf p}$\, of
the particle is rotated around the axis ${\bf e}_y$ ($i.e.$ the same as for the spin), and the rotation
angle is equal to $-2\theta<0$ (the rotation\, is in the counterclockwise direction and hence the negative sign).
It follows that the rotation of the spin ${\bf s}$\, relative to the direction of ${\bf p}$\,
is
\bq
\varphi_c^{UR} \simeq \pi - 2\theta - (-2\theta) =\pi.
\label{helfli}
\eq
It is easy to see that this implies that the projection of ${\bf s}$ onto the direction of
${\bf p}$ changes sign after the impact with the plane. This is equivalent to saying that the helicity of 
the particle changes sign after the impact. At first sight, this might appear counterintuitive if one takes 
the view that in the ultrarelativistic
limit the mass of the particle can be ignored, and thus one can assume that we deal with massless fermions. In
this case one might think that we could restrict to definite helicity states $h=\pm 1/2$, which would be in
contradiction with (\ref{helfli}). The argument however is not valid, if one recalls that for massless fermions such
states are equivalent to definite chirality states. The point is that the MIT boundary conditions (\ref{mitboc}) are
not chiral invariant \cite{chod},\, so that they are incompatible with a dynamics involving only well-defined
chiral states. The helicity flip (\ref{helfli}) has thus to be seen intimately connected with the massive nature of
the field.

\section*{5 Conclusions and discussions}

We considered a Dirac particle which bounces off from a perfectly plane described by chiral MIT
boundary conditions and determined the rotation of the spin after the reflection from the plane.
We made a comparison with the analogous result in the classical version of the problem, in which case the
reflecting plane was described by a repulsive scalar potential with translational invariance along
the plane. In the classical problem we also assumed that no external torque acts on the particle,
so that the evolution of the spin is completely determined by the Thomas precession effect. Under
these conditions, it turned out the rotation of the classical spin is independent on the
form of the repulsive potential.

We have found that (1) the rotation axes of the spin in the two descriptions coincide only for the chiral
angle $\phi =0$ (or, equivalently, $\phi = \pi$), (2) for these values of the chiral angle the
rotation angles coincide in the nonrelativistic limit, and (3) the two rotation angles also show a
remarkably good agreement up  velocities comparable to the speed of light. By contrast, (4) in the
ultrarelativistic limit the quantum rotation angle remains finite, while the classical angle diverges.

Points (1)-(3) are consistent with the fact that a chiral angle $\phi=0$ introduces
no specific interaction between the spin and the reflecting plane \cite{chod}, which at classical
level means that the spin evoluates as we assumed in the classical calculation. This supports the
general picture that the dynamics of the spin at the impact with a reflecting surface with $\phi=0$
is completely contained in the Thomas precession effect, and thus is\, of purely mechanical
nature. However, our result makes it clear that a complete quantum-classical similarity arises only
in the nonrelativistic limit. At sufficiently high velocities quantum effects become important
and the description based on the Thomas precession becomes inadequate.

Points (1) and (2) are somehow similar to the recovery of the nonrelativistic spin-orbit coupling from the
Dirac equation in an external electromagnetic field, in which case the result is given by the Thomas
term plus a magnetic interaction, where the last contribution arises due to the coupling
between the spin and the magnetic field in the proper frame of the particle. In our case no such
coupling exists, and thus only the Thomas term is determinant. It might be perhaps interesting to
see what type of interaction between the spin and the plane would assure the correspondence with
the quantum picture for $\phi \neq 0$. For example, a possible solution could be as in
ref.~\cite{jaffef} a delta-type magnetic field localized on the plane.

The rather unexpected result  of our calculation is point (4). The infinite difference between the
two rotation angles implies a serious discrepancy between the quantum and the classical picture in
the high velocity limit. In the rest of the section, we will make a few observations on this fact.

One can adopt essentially two views at this point. One is to simply admit that there is no reason for the spin of
a classical particle to reproduce the behavior of the spin of a quantum one, and thus there is no inconsistency
between the two results. This would then be just another instance in which the quantum and the classical theory
lead to  qualitatively different pictures ($e.g.$, one can think of the zero-point energy of a quantum and of 
a classical oscillator).

Another option would be to consider that the divergent angle in the classical picture is unphysical,\, and thus
something goes wrong in  the classical calculation ($e.g.$, one can think of the ultraviolet catastrophe in
the blackbody radiation problem). We incline\, to believe that this is the case. In the following, we will
comment on this possibility.

A frequent cause of an unphysical result in a theoretical model is an unrealistic idealization
in the model. In the case considered here, such an idealization can be identified in the fact
that the particle is assumed to behave as a perfectly rigid body. If one recalls the derivation of the Thomas
precession formula  (\ref{thopre}), this is implicit in the fact that the body is assumed from
the start to rigidly
rotate as a whole together with the Fermi-Walker transported axis of the particle's proper frame \cite{whee}.

A more realistic description would be to treat the classical particle as a finite size elastic body. It is then tempting
to conjecture that for such particles the rotation angle\, of the spin will\, stay finite in the
ultrarelativistic limit. Notice that elasticity effects will become important  precisely at large velocities
of the particle. Such velocities would imply large accelerations during the impact with the plane, which in turn
can lead to large deformations of the body. To prove this conjecture however is beyond the scope of our paper,
and most probably is not a simple task. Unfortunately, in order to solve this problem one should have at hand the
relativistic equations of motion for a deformable spinning body, which seems to be an unclarified issue
yet.\footnote{To our knowledge, the most suitable models that could be applied to our problem are the
quasi-rigid bodies\cite{ehle, stei3} in which the MPD equations are supplemented by quadrupole terms
induced by the spin and the velocity of the body. Using these models it was shown that the quadrupole effects
at large accelerations can significantly affect the trajectories \cite{bini1, bini2, bini3, bini4, bini5}.
See also the references in ref. \cite{stei3}, Sec. IV.}

We feel that an analogy might be relevant here. A long studied problem in classical electrodynamics is that of
the dynamics of charges in interaction with their proper field. It is well-known that  the equation of motion
for pointlike charges  (the Abraham-Lorentz-Dirac equation)  admits unphysical solutions, $i.e.$ acausal and
self-accelerated trajectories \cite{rohr, jack}. It is also well-known that the unphysical solutions can be
eliminated if one considers extended charges with a sufficiently large radius \cite{cald, moni1,levi, rohr1,
medi, grif, yagh}. Notably, the unphysical solutions can be also eliminated in the quantum theory,\footnote{In
the first quantized sense, as in our discussion. The solutions in question are the solutions of the Heisenberg
equation of motion for the position operator of the charge \cite{moni2} (in the nonrelativistic limit).} in which
case the associated Compton wave length acts as a sort of radius of the charge \cite{moni2}. It is appealing to
see the divergent rotation angle of the
spin obtained here as an  analogue of the unphysical trajectories for pointlike charges, with the perfect
rigidity of the body corresponding to the idealized limit of a zero radius of the charge. The same as retardation
effects in extended charges can eliminate the unphysical trajectories, one can hope that elasticity effects will
keep the rotation angle of the spin finite in the ultrarelativistic limit. It is also perhaps not a coincidence
that the divergent rotation angle arises in the high energy limit, which in the electrodynamical analogy would
correspond to the unphysical divergent  self-energy of pointlike charges.

Finally, a practical implication related to the above point could be the following. An astrophysical problem
that has received an increased interest in the last years is that of the trajectories of spinning bodies
in the vicinity of black holes.  When the mass of the black hole is much larger than that of the orbiting
body and the radius of the body is sufficiently small compared to the  local radius of curvature of the spacetime,
the motion  is well described by the MPD equations \cite{sing, saij}. We recall that the MPD equations are the
pole-dipole approximation in the multipole expansion of the Dixon's equations, and that this approximation is
justified  for a sufficiently small body, so that the internal tidal forces can be ignored \cite{dixo}.
Practically, this means that the MPD equations can be applied only to negligibly small, perfectly rigid bodies.
On the other hand, it is clear that for a  realistic body in sufficiently strong gravitational fields the tidal
forces can significantly deform the body, in which case the MPD will no longer apply.

Our observation is
that such a situation would  be similar to that in the classical problem discussed here in the ultrarelativistic
limit, in which case the large accelerations  during the impact with the reflecting plane would be the
analogue of the large accelerations in a strong gravitational field. The problematic divergent precession of
the spin in our calculation can then be suspected to occur also in the solutions of the MPD equations in strong
fields. One could further speculate from here that at very large velocities/accelerations the MPD equations
will generally $overestimate$ the precession of the spin compared to that of a finite size elastic object.
In other words, elasticity effects will tend to decrease the precession of the spin. It might be of interest
to see whether this phenomenon indeed occurs. For example, the effect might show up by comparing the predictions
of the MPD equations with numerical simulations of the evolution of physically realistic relativistic  bodies in
strong gravitational fields.\footnote{For example, there exists a wealth of work on numerical simulations for
spinning coalescing neutron stars or spinning neutron stars spiralling  into a black hole, see $e.g.$ refs.
\cite{tich, bern, taci,  maro, east}.} In a more unsophisticated setting, one could consider the models of the
quasi-rigid bodies mentioned above (see footnote 6) and examine the quadrupole effects in the precession of the
spin for accelerated bodies. Our conjecture then would be that at large velocities/accelerations the
quadrupole effects associated to the elasticity of the body will tend to decrease the precession compared to
that predicted by the standard Thomas precession formula (\ref{thopre}).

Finally,\, it could be relevant to mention that several recent works have noticed problematic solutions of the MPD
equations in the ultrarelativistic limit \cite{u1, u2, u3,u4, u5, u6}. However, they are concerned with unphysical
aspects in the orbital space (superluminal velocities or divergent accelerations), and not in the spin space.
Nevertheless, it might be interesting to see if the solutions adopted there could also be used to eliminate the
unphysical behavior in the spin space we noted here.

\section*{Acknowledgements}

I thank Professors Alexei Deriglazov and Alexander Silenko for pointing me to their works. I also thank 
my friends Ailedi Ivanovici and Attila Farkas for moral support during the writing of the paper.

\end{document}